\documentclass[prd,twocolumn,floatfix]{revtex4}

\newcommand{\be}{\begin{eqnarray}}
\newcommand{\ee}{\end{eqnarray}}

\newcommand{\vE}{{\bf E}}
\newcommand{\vS}{{\bf S}}

\newcommand{\va}{{\bf a}}

\newcommand{\ve}{{\bf e}}
\newcommand{\vf}{{\bf f}}

\newcommand{\vvr}{{\bf r}}
\newcommand{\vu}{{\bf u}}

\newcommand{\vx}{{\bf x}}

\newcommand{\ffp}{p}

\newcommand{\dS}{{\rm d}S}

\newcommand{\binomial}[2]{\left( \!\! \begin{array}{c} {#1}\\{#2} \end{array} \!\! \right) }

\newcommand{\dtau}{\rm d\tau}

\newcommand{\dby}[2]{ \frac{{\rm d} #1}{{\rm d} #2}}

\newcommand{\Grad}{\mbox{\boldmath $\nabla$}}
\newcommand{\Div}{\mbox{\boldmath $\nabla$} \cdot}

\usepackage{graphicx}
\usepackage{amssymb}

\begin{document}

\title{Self-force of a rigid ideal fluid, and a charged sphere in hyperbolic motion}

\author{Andrew M. Steane}
\affiliation{Department of Atomic and Laser Physics, Clarendon Laboratory, Parks Road, Oxford OX1 3PU, England.}


\begin{abstract}
We present two results in the treatment of self-force
of accelerating bodies. If the total force on an extended rigid object is calculated
from the change of momentum summed over planes of simultaneity of successive rest
frames, then we show that an ideal fluid, moving rigidly, exerts no net force on its boundary.
Under this same definition of total force, we find
the electromagnetic self-force for a spherical charged shell of proper
radius $R$ accelerating with constant proper acceleration $g$ is
$(2 e^2 g/R)[1/12 - \sum_{n=0}^\infty (g R)^{2n} ((2n-3)(2n-1)(2n+1)^2)^{-1}]$.
\end{abstract}

\maketitle

Recent work on self-force and radiation reaction in classical electromagnetism has clarified
various issues which had been unclear for a considerable period (roughly one hundred years).
Several of the difficulties that persisted for a long time were concerned with taking the
point-limit for an entity possessing finite charge and observed mass. These problems go away
when one insists that there is no such limit, because it is not legitimate to assume the
non-electromagnetic contribution to the mass is negative
\cite{14SteaneB,61Erber,14Burton,91Ford,09Gralla}. One then has to grapple with
the fact that an exact treatment of motion of charged entities cannot deal with point-particles,
if the charge is not itself infinitesimal, but must treat extended objects or distributions
of charge. 

As soon as one has to deal with an extended distribution of charge, the question of its
field and motion is no longer a question of finding a single worldline and applying standard
formulae to find the field; instead one has to find the shape of a world-tube, and the self-interaction
cannot in general be calculated exactly in closed form. 
Owing to the complexity of the problem, there are few 
exact general statements than can be made, either about the forces on, or the
electromagnetic fields produced by, accelerating charged bodies. 
In this situation it is useful to identify some statements
that have as wide an application as possible, and to obtain some example
non-trivial exact solutions. 

A modest but useful step of the first kind was taken by
Ori and Rosenthal \cite{03Ori}, who pointed out that the summed electromagnetic interaction
of a pair of point charges of fixed proper separation was independent
of the orientation of the line between the charges, as long as it is summed a certain way
that we describe in section \ref{s.def}. A step of the second kind was
taken in \cite{13Steane}, where electromagnetic self-force was calculated exactly
for a given case.
In the present paper we obtain a further general statement about rigid-body motion,
namely: the total force exerted by an ideal fluid on its boundary, when the fluid
moves rigidly, is zero, for a certain natural definition of this force, namely that adopted
in \cite{03Ori,64Nodvik,82Pearle}.
We then extend the
result of \cite{13Steane} by obtaining the electromagnetic self-force of a 
charged spherical shell in rigid hyperbolic motion, taking the sum over the body in
the same sense as recommended in \cite{03Ori,64Nodvik,82Pearle}.
We also invoke the results of \cite{14SteaneA} to deduce
that the result is applicable to a charged spherical shell at rest in a uniformly accelerating
reference frame (the Rindler frame) and observed by an observer at rest in that frame.

\section{Defining the self-force}  \label{s.def}

In relativistic calculations concerning extended bodies,
the first issue that arises is, how to define properties such as energy and momentum. This
is non-trivial because when the entity is not isolated, the total four-momentum of its parts
can depend on the choice of spacelike hypersurface over which they are summed. This
dependence does not arise for an isolated body \cite{12Steane}, but a charged
body is never isolated because it
is in permanent interaction with its own electromagnetic field. 

This issue was discussed in \cite{64Nodvik,82Pearle,03Ori} and somewhat more generally
in \cite{14SteaneA}. Consider a composite object that can be decomposed into a set of discrete entities $i$.
The total 4-momentum of the composite object may be defined as the sum of the 4-momenta of its
parts, where the sum is taken over some spacelike hypersurface which we denote by $\chi$:
\be
p^\mu_{\rm tot}(\tau_c,\chi) = \sum_{i}  p_i^\mu \left(\tau_{i,\chi}\right) . \label{sump}
\ee
$\tau_{i,\chi}$ is the proper time on the $i$'th worldline when that worldline intersects $\chi$,
and $\tau_c$ is the proper time on some reference worldline (e.g. the worldline of the centroid).
Typically, one picks a spacelike hyperplane (so that the events $\{i\}_\chi$ are
simultaneous in some frame). In general, $p^\mu_{\rm tot}$ depends on $\chi$
so this `total 4-momentum' is not a property of the object alone, and it will not behave
like a 4-vector unless we specify $\chi$ in a suitably covariant way. A suitable way, is, for example,
to pick $\chi$ such that it is the same surface, irrespective of what frame may be used to
calculate the four components of each $p_i$.

For an object whose motion is rigid---that is, its motion is such that at any given event on
the world-tube there is a reference frame in which all parts of
the object are at rest, and at the same proper distances---a natural choice of~$\chi$ is
the hyperplane of simultaneity for the instantaneous rest frame (IRF) at the given~$\tau_c$.
By making this choice we obtain a well-defined 4-vector $p^\mu_{\rm tot}$, but it
is not the only possible choice. {\em Any} recipe that picks out a unique hyperplane $\chi$
for each $\tau_c$ will result in a $p^\mu_{\rm tot}$ that transforms in the right way,
because the terms in the sum on the right hand side of (\ref{sump}) all do, and the
recipe fixes the set of events in a frame-independent way. As a result, when calculating
the components of  $p^\mu_{\rm tot}$ relative to any given inertial frame at some given
$\tau_c$, one may be summing
over events that are not simultaneous in that frame, but one must accept this in order
to have a well-defined 4-vector.

If there are several bodies in different states of motion in a given problem,
there will not be a unique IRF for all of them, so the policy of adopting the
IRF in order to define $\chi$ is not necessarily the only policy
that makes sense.

Once we have decided how to choose $\chi$ for each $\tau_c$, it becomes possible
to define the total rate of change of 4-momentum:
\be
\dby{\ffp^\mu_{\rm tot}}{\tau_c} = \lim_{\delta\tau_c \rightarrow 0}
\frac{\ffp^\mu_{\rm tot}(\tau_c + \delta\tau_c, \chi + \delta\chi) - \ffp^\mu_{\rm tot}(\tau_c,\chi)}{\delta\tau_c}
\label{defforce}
\ee
where we have assumed a one-to-one correspondence between $\chi$ and $\tau_c$, such that 
$\delta\chi \rightarrow 0$ as $\delta\tau_c \rightarrow 0$. Hence \cite{03Ori,14SteaneA}
\be
\dby{\ffp^\mu_{\rm tot}}{\tau_c} &=& 
 \sum_i \dby{\ffp^\mu_i}{\tau_i}  \dby{\tau_i}{\tau_c}   \label{ptot}
\ee
where each ${\rm d}\tau_i$ is the proper time elapsed
on the $i$'th
wordline between the intersections of that worldine with $\chi$ and $\chi+{\rm d}\chi$,
and  the quantities ${\rm d}\ffp^\mu_i/\dtau_c$
and $\dtau_i / \dtau_c$ are evaluated on the hyperplane $\chi$. 

In \cite{13Steane} the case of a rigid spherical shell was considered, with $\chi$ taken
as the hyperplane of simultaneity of the IRF at some instant,
and $\chi+{\rm d}\chi$ taken as a parallel hyperplane such that $\dtau_i = \dtau_c$.
This is a legitimate choice, as we made clear above, but it is not the only one that
makes sense, and another case is interesting, namely, to take for
$\chi+{\rm d}\chi$ the new hyperplane of simultaneity, i.e. the one associated
with the next state of motion, which is not the same as the initial one when the
body is accelerating. This is the choice made by Nodvik \cite{64Nodvik} and also
recommended in \cite{03Ori,82Pearle}. Therefore we will explore its impact on
the case of the sphere in rigid hyperbolic motion in section \ref{s.em}.
First we apply it to obtain a useful observation concerning 
internal stress.

\section{Self-force of a rigid ideal fluid}

We consider the net force on a rigid body owing to its internal stress. 

Rigid motion as defined above (at each moment there is an IRF for the whole body, 
and the proper shape and size does not change) implies motion in a straight line,
but the acceleration need not be constant. Such motion is treated in section II
of \cite{64Nodvik} and section 9.2.1 of \cite{12Steane}. The condition for rigidity
is that worldlines at points A and B on the body should satisfy Eqs (9.23) and (9.24)
of \cite{12Steane}, which we reproduce here for convenience:
\be
x_B - x_A = \gamma L_0, \;\;\;\; t_B - t_A = \gamma v L_0
\ee
where $x_A,t_A,x_B,t_B$ are coordinates in an inertial frame, $L_0$ is the
proper separation of $x_B$ and $x_A$, $v = d x_A/d t_A$ and
$\gamma = (1-v^2)^{-1/2}$ (taking $c=1$). From these equations we find
\be
\dby{^2 x_B}{t^2_B} = \frac{\dot{v}}{1 + \dot{v} L_0}   \label{d2xb}
\ee
where $\dot{v}$ is the proper acceleration on worldline A. Now, by definition,
for rigid motion, there is always an IRF. If we evaluate (\ref{d2xb})
in this frame, then the quantity on the left hand side is
the proper acceleration on worldline B, and $L_0 = x_B-x_A$.
Then we have
\be
\dby{^2 x_B}{t^2_B} = \frac{x_A - x_0}{x_B - x_0} \dot{v}  \label{dxB}
\ee
where $x_0 = x_A - 1/\dot{v}$. Hence we have that for
rigid motion in the $x$ direction, there exists a
plane $x=x_0$ in the IRF, such that the acceleration of each part
of the body is inversely proportional to its distance from
the plane.

Now suppose the interior of the body in question behaves as an ideal fluid,
i.e. it can support pressure or tension but not sheer stress. For such
a fluid, in the IRF the Navier-Stokes equation takes the form
\be
(\rho_0 + p) \frac{D\vu}{Dt} = -\Grad p
\ee
where $p$ is the pressure and $\rho_0$ the mass density. Choose
the origin such that $x_0 = 0$, then, using (\ref{dxB}), the acceleration $D\vu/Dt \propto
\hat{\vx}/x$ and therefore the pressure satisfies the differential equation
\be
\dby{p}{x} = - \frac{\rho_0 + p}{x}  \label{dpdx}
\ee
whose solution is
\be
p = \frac{\rm const}{x} - \rho_0.
\ee

The total force exerted by the ideal fluid on its boundary is, using (\ref{ptot}),
\be
\vf_p = \oint \dby{\tau_x}{\tau_c} p \,{\rm d}\vS .
\ee
We now make our choice of the hyperplanes $\chi$ and $\chi + {\rm d}\chi$
in order to evaluate $\dtau_x/\dtau_c$. We adopt the hyperplanes of simultaneity of
the two successive IRFs, which leads to
\be
\dby{\tau_x}{\tau_c} = \dot{v} x        \label{vdotx}
\ee
where $\dot{v}$ is the proper acceleration at $x=x_c$. The proof of this
equation is given in the next section. Assuming it for the present argument, we have
\be
\vf_p = \oint \dot{v} x p \,{\rm d}\vS .
\ee
Let $\ve_i$ be a unit vector in the $i$'th direction. Then the component of
the force in the $i$'th direction is
\be
\vf_p \cdot \ve_i = \oint \dot{v} x p \ve_i \cdot {\rm d}\vS 
= \int \Div (\dot{v} x p \ve_i) {\rm d}V.
\ee
Now,
\be
\Div (\dot{v}xp\ve_i) = \dot{v} \left(p+x \dby{p}{x}\right) \delta_{1i} = 
-\dot{v} \rho_0 \delta_{1i}
\ee
using (\ref{dpdx}). Hence
\be
\vf_p = - m_0 \va           \label{fp}
\ee
where $\va$ is the proper acceleration on the reference worldline, and
 $m_0 = \int \rho_0 {\rm d}V$ is the `completely bare' mass, i.e. the mass
of the fluid without taking either the stress or any electromagnetic contribution into
account. Hence
we have that, {\em for rigid motion of an ideal fluid, the pressure or tension in
the fluid gives no net force on its boundary, when it is defined as the rate of change
of total momentum evaluated in the sequence of IRFs.}
Alternatively, the same result may stated as that the
contribution to the intertial mass that we might have expected the stress to
give, namely $\int p \,{\rm d}V$, is exactly balanced by the difference in pressure
across the body.

We have not needed to make any assumption about the shape of the body
in order to derive equation (\ref{fp}). It applies equally to a dipole as to a sphere,
for example. In particular, it shows that the net force owing to internal pressure
does not depend on the orientation of the body relative to its acceleration, when
it is evaluated using the sequence of IRFs to define
the hyperplanes on which the forces are summed
(the forces being conveniently summed by appropriate use of Eq. (\ref{ptot})). This gives a further reason,
in addition to the one obtained in \cite{03Ori}, to recommend this choice
when discussing self-force for bodies moving rigidly.

\section{Electromagnetic self-force of a rigid spherical shell in hyperbolic motion} \label{s.em}

The electromagnetic self-force of a uniformly accelerating rigid charged spherical shell was calculated
in \cite{13Steane}. That calculation gave the result of summing the electromagnetic force over
such a sphere in its instantaneous rest frame. The force was taken to be given by the sum of the local forces
on each element of charge, evaluated at events simultaneous in the rest frame. In our present notation,
this amounts to taking $\chi+{\rm d}\chi$ parallel to $\chi$ and hence
$\dtau_i = \dtau_c$ in Eq. (\ref{ptot}).  This is a legitimate choice, but not necessarily the most
useful one. As we have seen in the previous section, and also for the reasons given in \cite{03Ori,64Nodvik},
it is useful to consider also the case where $\chi+{\rm d}\chi$ is the hyperplane of simultaneity of the
next IRF. This is the subject of this section.

An object undergoing rigid hyperbolic motion in the $x$-direction
will be at rest relative to the constantly accelerating reference
frame described by the Rindler metric:
\be
g_{ab} = {\rm diag}(-x^2,1,1,1).
\ee
The line element is
\be
{\rm d}s^2 = - x^2 {\rm d}t^2 + {\rm d}x^2 + {\rm d}y^2 + {\rm d}z^2  . \label{xmetric}
\ee
For this metric, the local `acceleration due to gravity' (that is, the acceleration, relative to the
coordinate frame, of an object in free fall) is $-1/x$.

Lines at constant $(x,y,z)$ are wordlines of points on a body at rest in the frame. Hyperplanes at constant $t$
are planes of simultaneity for the successive IRFs. The proper time between events on two
such planes at any given $(x,y,z)$ is given by
\be
\dtau_x = x \,{\rm d}t.
\ee
Therefore 
\be
\dby{\tau_x}{\tau_c} = \frac{x}{x_c} = g x,      \label{gx}
\ee
where $\tau_c$ is proper time along a reference worldline which we take as the one at
$(x,y,z) = (x_c,0,0)$, and $g=1/x_c$ is the proper acceleration on this reference worldline.
Hence for this case, Eq. (\ref{ptot}) reads
\be
\dby{p^\mu_{\rm tot}}{\tau_c} 
= g \sum_i  \dby{p^\mu_i}{\tau_i} x_i    .   \label{ptotx}
\ee
The same argument can be used to obtain the slightly more general Eq. (\ref{vdotx}), in which
$\dot{v}$ is not necessarily constant. This is done by adopting the Rindler frame matched to
the instantaneous value of $\dot{v}$, for a body whose acceleration can change.

We now apply this to find the electromagnetic contribution to the self-force of a spherical
shell of charge undergoing rigid hyperbolic motion. This self-force is given by summing
the force ${\rm d}\vf$ on each element of charge ${\rm d}q$ making up the shell, where
${\rm d}\vf$ is owing to the electromagnetic field sourced by the rest of the shell. 
The calculation follows the argument of \cite{13Steane} closely, so we shall refer to that paper as
PRS, and its equations as (PRS$n$) where $n$ is the equation number.

In the case under consideration, the charge is concentrated in a thin spherical shell. Therefore,
by the argument leading to Eq. (PRS1.4)  we have
\be
{\vf}_{\rm self} = \oint g x \sigma \frac{\vE_- + \vE_+}{2}  {\rm d}S
\ee
where $\vE_- = \lim_{\epsilon \rightarrow 0} \vE(R-|\epsilon|)$ is the field on the interior surface of the shell, and
$\vE_+ = \lim_{\epsilon \rightarrow 0} \vE^(R+|\epsilon|)$ is the field on the exterior surface of the shell. $R$
is the radius of the shell.
By standard reasoning from Maxwell's equations in an inertial frame we have
$\vE_+ = \vE_- + (\sigma/\epsilon_0) \hat{\vvr},$
therefore
\be
{\bf f}_{\rm self} = \sigma   \oint g x \vE_- \dS + \vf_\sigma    \label{finteg}
\ee
where
\be
\vf_\sigma &=& \oint \frac{\sigma^2}{2 \epsilon_0} \frac{\vvr}{R} g x \dS
\;=\; \frac{e^2 g}{6 R} \hat{\vx}
\ee
where $e^2 = q^2/4\pi\epsilon_0$ (in SI units) if $q$ is the total charge of the shell,
and $\vvr$ is the vector $(x-x_c,\,y,\,z)$.

To calculate the integral in (\ref{finteg}), note that only the $x$-component is non-zero, and use
\be
E_{-,x}(x,\rho) = \sum_{n=0}^\infty a_n E_{x,n}(x,\rho)
\ee
where $E_{x,n}(x,\rho)$ is given by equation (PRS4.17) after suitably scaling the units,
and the calculation of the
coefficients $a_n$ is described in PRS. We find
\be
f_{{\rm self},n} &=& 2 \pi R \sigma \int_{(1/g)-R}^{(1/g)+R} g x E_{x,n}(x,\rho(x)) {\rm d}x \\
&=& \frac{2 g e^2}{R} \left\{
(-1)^{n+1} +  \sum_{m=1}^n \frac{(-1)^{m+n}\binomial{n}{m}  (gR)^{2m}}{(2m-1)(2m+1)}
 \right\}.  \nonumber
\ee
The self-force is then given by
\be
f_{\rm self} = f_\sigma   +  \sum_{n=0}^\infty a_n f_{{\rm self},n}           \label{fselfsum}
\ee 
The end result is that the electromagnetic self-force of the uniformly charged spherical shell undergoing
rigid hyperbolic motion with proper acceleration $g$ is
\be
f_{\rm self} = \frac{g e^2}{R}\left( \frac{1}{6} - 2 \sum_{n=0}^\infty \frac{(g R)^{2n}}
{(2n-3)(2n-1)(2n+1)^2} \right),   \label{fselfexact}
\ee
where I have calculated this result for terms up to order $R^{101}$, and I conjecture
its validity at all orders. The result differs from (PRS5.11) because of the different choice of
hyperplane $\chi+{\rm d}\chi$, as we have explained above.
At $g=1$ and $e^2=1$ the first few terms in this series are
\be
f_{\rm self} \simeq \frac{-1}{2 R} + \frac{2}{9}R - \frac{2}{75} R^3 +
\frac{2}{735} R^5 + \frac{2}{2835} R^7 +\cdots .   \label{fseries}
\ee
Eq. (\ref{fseries}) agrees with Nodvik \cite{64Nodvik}, who considered a shell 
undergoing arbitrary motion and
calculated the lowest order terms in the series expansion. For the case of hyperbolic motion
Nodvik's result (Eqs (7.21)-(7.24) of \cite{64Nodvik})  is
\be
f_{\rm self}^{\rm Nodvik} = -\frac{1}{2R} + \frac{2}{9} R + O(R^3),  \label{Nodvik}
\ee
Nodvik formulated the problem in such a way that he calculated the self-force
using a prescription that corresponds to our Eq. (\ref{ptotx}), so we expect his
result to agree with ours, as it does. The reason for the disagreement with (PRS5.11) was wrongly
described in (\cite{13Steane}), where it was attributed to the effect of internal stress.
We can now understand this better through the insight described in the previous section, leading
to Eq. (\ref{fp}). When the momenta are summed in successive IRFs, a calculation which
neglects to include the effects of internal stress will nevertheless arrive at a 4-vector force with
the correct Lorentz transformation properties, because the pressure does not appear in Eq. (\ref{fp}).
In short, the `4/3 problem' does not arise. But this does not mean that it is correct to neglect
the stress. In fact, only when it is correctly included does one arrive at Eq. (\ref{fp}).

\subsection{Electromagnetic self-force in the Rindler frame}

So far we have only discussed forces observed by inertial observers. We made use of
the Rindler metric in order to obtain Eq. (\ref{gx}), but (\ref{fselfexact}) is the self-force
observed by an inertial observer, relative to whom the sphere accelerates.
We now briefly comment on the electromagnetic self-force observed by an observer
at rest relative to the sphere; that is, an accelerated observer. Such an observer's
measurements of space and time are indicated by the coordinates $(t,x,y,z)$ appearing
in the line element (\ref{xmetric}), and therefore his most natural definition of
total force on an extended object at rest relative to him is the one given by
(\ref{ptotx}).

If he so chooses, inertial forces may be considered by the accelerated observer as an
example of a gravitational field. Such an observer finds that the field-lines of the sphere
`droop' in the gravitational field, and consequently the sphere exerts once again a self-force,
owing to the fact that the electric forces between all the pairs of charges composing the
sphere do not sum to zero. It was shown in \cite{14SteaneA} that the electromagnetic force
observed by such an accelerated observer is the same as the one observed by the inertial
observer who is at rest in the IRF of the sphere. Consequently it is given
by (\ref{fselfexact}).

To conclude, the main results of this paper are Eqs (\ref{fp}) and (\ref{fselfexact}).
(\ref{fp}) is a general statement about rigid motion of a body whose internal forces
correspond to those of an ideal fluid. The shape and the variation of the acceleration with
time are arbitrary. Eq. (\ref{fselfexact}) is a correct to high order, and conjectured exact,
statement of the electromagnetic self-force of a rigid spherical shell undergoing hyperbolic
motion. It is a partner to Eq. (5.11) of \cite{13Steane}. The two equations differ because
there is no unique definition of `the' self-force, owing to the dependence on the choice of
hypersurfaces in spacetime over which the momenta are summed. However, the choice made in
the present paper is arguably the more neat or natural. The previous study 
enabled the fields to be found exactly, no matter how the total force is defined, and
gave the self-force under another reasonably natural definition.

\bibliography{selfforcerefs}

\end{document}